\documentclass[runningheads]{llncs}

\usepackage[T1]{fontenc}
\usepackage{graphicx}
\usepackage{amssymb}
\usepackage{booktabs}
\usepackage{placeins}

\begin{document}

\title{ASR Personalization in Pathological Voice: Accuracy, Efficiency, and Duration Effects}
\titlerunning{Personalised ASR for Pathological Speech}

\author{
Giuseppe Giaimis\orcidID{0009-0005-9383-2109}
\and Sergei Katkov\orcidID{0009-0006-1000-295X}
\and Muhammad Azfar Yaqub\orcidID{0000-0003-2150-952X}
\and Antonio Liotta\orcidID{0000-0002-2773-4421}
}

\authorrunning{G. Giaimis et al.}

\institute{
Free University of Bolzano, Bolzano, Italy\\
\email{\{GGiaimis, Sergei.Katkov, MYaqub, Antonio.Liotta\}@unibz.it}
}

\maketitle

\begin{abstract}
Automatic Speech recognition (ASR) struggles with Pathological speech due to persistent speaker-specific mismatch. In clinical and accessibility scenarios, only limited target-speaker data is typically available, making full model retraining impractical. We present a component-level analysis of LoRA adaptation in Whisper, investigating how adapter placement interacts with the model size, training efficiency and utterance-duration in single-speaker settings. Our experimental comparative study shows that while scaling reduces errors, it does not eliminate mismatch. Furthermore, the full attention LoRa provides the most reliable accuracy, and the best time--accuracy trade-off. 
Improvements vary across utterance length, with medium-length recordings benefiting the most, while short and long utterances remain challenging. These findings highlight the importance of adaptation placement and duration-dependent evaluation, under challenging, data-scarce personalization scenarios. 
\end{abstract}


\section{Introduction}
\label{sec:introduction}

Modern ASR systems are primarily built and evaluated on typical speech. Yet, speakers suffering from voice pathologies (i.e., atypical phonation, articulation, prosody, and reduced intelligibility) lead to errors that severely affect communication and access to AI-assisted digital services. The errors are not uniform: short commands, longer utterances, and persistent speaker-specific acoustic patterns can fail in different ways.

Whisper is a strong general-purpose ASR model trained on 680{,}000 hours of weakly supervised multilingual speech~\cite{radford2023robust}. However, robustness studies show that average performance can hide failures under distribution shift~\cite{evaluating_asr,katkov2026robustness}. Recent investigations of Whisper and other ASR systems under diverse audio transformations and challenging acoustic conditions further confirm substantial degradation under non-standard speech scenarios~\cite{benchmarking_whisper,asr_systems_under_acoustic_challenges}.Pathological speech is such a shift, as it reflects a persistent speaker-specific property rather than a transient noise condition. While scaling the base model size tends to reduce errors, clinical and accessibility settings require speaker-specific adaptation from very limited amount of target-speaker data.

Parameter--efficient methods such as LoRA are well suited to this setting, as they freeze the base model and store only a small set of speaker-specific weights. However, for encoder-decoder ASR models, it remains unclear where these weights should be placed: on the decoder, the encoder, or both. Furthermore, it is important to understand whether the most accurate placement is efficient and whether its gains are stable across utterance lengths.
To address these questions, we investigate the following research questions:
\begin{enumerate}
    \item How does a zero-shot Whisper perform on a pathological speaker scale with model size?
    \item Which attention-side LoRA placement gives the best fixed-rank adaptation (decoder-only, encoder only, or full-attention)?
    \item Does the most accurate placement also offer the best time--accuracy trade-off?
    \item How does adaptation performance vary across short, medium, and long utterances?
\end{enumerate}

We address these questions through a controlled ablation study, developed around a real  pathological speaker, to exemplify ASR applications and methodologies in precision medical context. Four Whisper model sizes (tiny, base, small, medium) are evaluated across three LoRA placements under a fixed configuration (rank, learning rate, dropout, and leave-one-out protocol).  The design isolates the effect of adapter placement under data-scarce conditions. 

Our findings show that decoder-side adaptation is often ineffective, or even detrimental, while full-attention LoRA provides the most consistent accuracy and the best observed time--accuracy trade-off. Moreover, duration tiers analysis reveals that improvements are not uniform; medium-length recordings benefit the most, while short and long recordings expose persistent errors.
Overall, this study pinpoints hard challenges in ASR-based precision medicine, which are the basis of a more inclusive access of patients to AI-assisted technologies.


\section{Related Work}
\label{sec:related_work}

Modern ASR systems perform well for typical speakers in standard recording conditions, but their accuracy drops severely when the speech signal deviates from the patterns seen during large-scale pretraining. Pathological and dysarthric speech create this kind of mismatch: atypical phonation, articulation, prosody, and intelligibility, which are persistent speaker-intrinsic properties rather than transient environmental distortions.

Recent surveys describe progress from acoustic-model adaptation toward deep learning, self-supervised representations, and multimodal systems; yet data scarcity and high speaker variability remain
key limitations~\cite{bhat2025dysarthric,qian2023dysarthric}. These characteristics make speaker-specific personalisation especially relevant: limited target-speaker data has been shown to improve recognition for dysarthric, accented, and disordered speech~\cite{green2021personalized,shor2019personalizing}, and atypical speech has also been addressed using residual adapters and speaker-conditioned hypernetworks~\cite{muller2024hypernetworks,tomanek2021residual}.

Large pretrained systems such as Whisper, wav2vec~2.0, and Conformer establish strong ASR baselines~\cite{baevski2020wav2vec,gulati2020conformer,radford2023robust}; yet strong average performance does not guarantee robustness under distribution shift. Previous work has shown that modern ASR systems degrade substantially under adverse acoustic conditions. In particular, Katkov et al.~\cite{benchmarking_whisper} demonstrated significant variations in Whisper performance under diverse audio transformations and real-time constraints, while a multilingual evaluation revealed similar vulnerabilities across multiple ASR architectures and acoustic challenges~\cite{asr_systems_under_acoustic_challenges}. Furthermore, robustness analyses under synthetic transformations and real noise conditions have shown that strong average performance can mask substantial failures under distribution shift~\cite{katkov2026robustness}. Similar findings have been reported in multilingual evaluations under adverse acoustic conditions, where performance varies substantially across languages and recording environments~\cite{asr_systems_under_acoustic_challenges}. However, such shifts are mostly external to the speaker, whereas pathological voice quality is persistent across utterances and therefore motivates
speaker-specific adaptation rather than noise-only robustness.

To address these challenges, parameter-efficient fine-tuning (PEFT) offers a practical alternative to full model updates in data-scarce personalisation settings. LoRA keeps the pretrained model frozen and adds trainable low-rank updates to selected weight matrices~\cite{hu2022lora}; it has also been applied directly to Whisper ASR~\cite{song2024lorawhisper}. Related PEFT methods, including BitFit and other compact adaptation strategies, show that small targeted updates can be effective ~\cite{benzaken2022bitfit,he2022unified}.

For an encoder-decoder model such as Whisper, however, it remains unclear where such updates should be applied. The encoder is a plausible target because it processes the acoustic representation, and prior work suggests that Whisper encoders retain acoustic information beyond transcription~\cite{gong2023whisperat}. The decoder, in contrast, controls autoregressive generation and
cross-attention to encoder states. This motivates our component-level comparison of decoder-only, encoder-only, and full-attention LoRA under a shared fixed-rank training configuration.

\section{Methodology}
\label{sec:methodology}

\subsection{Dataset}
\label{sec:dataset}

The dataset consists of 215 audio recordings of a single Italian female speaker with a medical condition, provided in WAV format at 16~kHz. The total duration is 1{,}321 seconds ($\approx$22 minutes). The average length of a recording is 6.15 seconds, ranging from 0.48~s and 26.37~s.

All recordings are transcribed in Italian, with utterances containing an average of 9.9 words (range: 3--54). For analysis, the data are grouped into three duration tiers (Table~\ref{tab:duration_tier}) used for downstream analysis. The thresholds are empirical bins that distinguish short commands, medium-length recordings (the dominant  category), and longer spoken responses. Metadata on pathology subtype, severity, detailed demographics, and recording condition are not available, due to stringent ethical constraints. 


\begin{table}[h]
    \centering
    \caption{Duration-tier distribution of the dataset.}
    \label{tab:duration_tier}
    \begin{tabular*}{0.80\textwidth}{@{\extracolsep{\fill}}lccc@{}}
        \toprule
        \textbf{Tier} & \textbf{Threshold} & \textbf{N} & \textbf{Avg.\ duration} \\
        \midrule
        Short  & $< 3.5$~s   & 21  & 2.50~s  \\
        Medium & 3.5--6.0~s  & 139 & 4.81~s  \\
        Long   & $> 6.0$~s   & 55  & 10.81~s \\
        \bottomrule
    \end{tabular*}
\end{table}

\subsection{Acoustic Characterisation of the Target Speaker}
\label{sec:acoustic_characterisation}

To characterise the acoustic properties of the target speaker, we computed a non-diagnostic acoustic analysis over all 215 recordings. Because the recordings consist of short prompted phrases, we used smoothed cepstral peak prominence (CPPS) as the primary descriptor, as cepstral measures are recommended for instrumental voice assessment and are robust for running speech~\cite{hemanackah2003cpp,patel2018protocols,murton2020cpp}. Additionally, Harmonics-to-Noise Ratio, jitter, and shimmer are included as supportive descriptors. These measures are more sensitive to reliable voicing and cycle extraction in connected speech~\cite{brockmann2011jitter,teixeira2013vocal,zhang2008running}. For comparison, we used 500 randomly selected female samples from Common Voice~24, processed with the same extraction pipeline.

\begin{table}[h]
    \centering
    \caption{Acoustic characterisation of the target speaker and comparison
    with a female Common Voice (CV) reference subset processed with the same
    extraction pipeline.}
    \label{tab:acoustic_characterisation}
    \small
    \begin{tabular}{lccc}
        \toprule
        \textbf{Metric} & \textbf{Target median} & \textbf{CV median} & \textbf{Reference} \\
        \midrule
        CPPS (dB) & 4.69 & 7.94 &
        $>6.49$~dB~\cite{buckley2023normative} \\
        HNR (dB) & 1.32 & 12.20 &
        $6.58$--$12.22$~dB~\cite{tyleckova2019connected} \\
        Jitter local (\%) & 4.48 & 2.12 &
        $0.72$--$2.12\%$~\cite{tyleckova2019connected} \\
        Shimmer local (\%) & 18.04 & 9.85 &
        $8.33$--$16.81\%$~\cite{tyleckova2019connected} \\
        \bottomrule
    \end{tabular}
\end{table}

Table~\ref{tab:acoustic_characterisation} shows a consistent separation between the target speaker and the female Common Voice reference subset. Specifically, the CPPS median for the target lies below the retained reference threshold, whereas the Common Voice median exceeds it. Similarly, HNR is lower for the target speaker, while jitter and shimmer are higher, with all three measures deviating from the connected-speech reference ranges shown in the table. Together these measures provide descriptive evidence consistent with dysphonic/pathological voice characteristics, although they do not constitute a clinical diagnosis.

\subsection{Whisper Models}
\label{sec:models}

Whisper is an encoder-decoder transformer architecture~\cite{radford2023robust}. The encoder begins with two \texttt{Conv1d} layers applied to the 80-channel log-mel spectrogram, followed by transformer blocks that process the resulting sequence. This ordering is important because the convolutional front-end processes the acoustic signal before any attention layer. LoRA modifies only selected attention projections (\texttt{q\_proj} and \texttt{v\_proj})~\cite{hu2022lora}, while all other parameters remain frozen in our experiments. As a result, the convolutional front-end lies outside the adaptation path. Table~\ref{tab:whisper_models} lists the model sizes used in this study and reports the size of the convolutional encoder front-end, which remains unchanged across all configurations.

\begin{table}[h]
    \centering
    \caption{Whisper variants used in this work. CNN params denotes the
    parameters in the two \texttt{Conv1d} layers of the encoder front-end.}
    \label{tab:whisper_models}
    \small
    \begin{tabular*}{0.98\textwidth}{@{\extracolsep{\fill}}lrrrrrr@{}}
        \toprule
        \textbf{Model} & \textbf{Params} & \textbf{Enc.} & \textbf{Dec.} & \textbf{Dim} & \textbf{Heads} & \textbf{CNN params} \\
        \midrule
        Whisper-tiny   & 39M  & 4  & 4  & 384  & 6  & 535{,}296 \\
        Whisper-base   & 74M  & 6  & 6  & 512  & 8  & 910{,}336 \\
        Whisper-small  & 244M & 12 & 12 & 768  & 12 & 1{,}955{,}328 \\
        Whisper-medium & 769M & 24 & 24 & 1024 & 16 & 3{,}393{,}536 \\
        \bottomrule
    \end{tabular*}
\end{table}

\subsection{Fine-tuning Strategies}
\label{sec:strategies}

We compare four conditions for each Whisper model size: one zero-shot baseline and three fixed-rank LoRA configurations. The goal is not to tune a separate configuration for each model, but to probe where a small speaker-specific update is most effective.

\textbf{Vanilla.}
The vanilla condition uses the original Whisper checkpoint without any parameter update. It serves as a baseline for how much of the speaker-specific mismatch is reduced by model capacity alone.

\textbf{Decoder-only LoRA.}
This configuration inserts LoRA modules only into decoder attention projections. It tests whether personalisation can be achieved mainly by changing the output-side pathway, which conditions on previously generated text and on encoder states. Weak performance in this setting suggests that the mismatch cannot be fully addressed by output-side adaptation alone.

\textbf{Encoder-only LoRA.} 
This configuration inserts LoRA modules only into encoder attention projections. It tests whether a small update after the convolutional front-end can improve the acoustic representation of the pathological speaker while leaving the decoder unchanged.

\textbf{Full-attention LoRA.}
This configuration applies LoRA to both encoder and decoder attention projections. It is the strongest parameter-efficient setting considered here, allowing adaptation on both sides of the sequence-to-sequence model; but it still freezes the convolutional front-end, feed-forward layers, normalisation layers, embeddings, and output projection. All LoRA variants modify the query and the value projections (\texttt{q\_proj} and \texttt{v\_proj}), with all other parameters kept frozen.

\subsection{Experimental Design}
\label{sec:design}

We evaluate these configurations in a controlled experimental design that crosses four Whisper sizes (tiny, base, small, medium) with three LoRA placements, in addition to the corresponding zero-shot baselines. This defines two main dimensions of analysis: model capacity and adapter placement. Because LoRA rank and optimisation settings are fixed across all adapted conditions, differences across decoder-only, encoder-only, and full-attention LoRA can be interpreted as placement effects rather than model-specific tuning.

In addition to overall accuracy, we include two practical analyses. First, an efficiency analysis compares WER reduction with average training time per fold, since personalisation should not require prohibitively slow speaker-specific updates. Second, a duration-tier analysis checks robustness across utterance lengths: short utterances provide limited context, while longer ones allow errors to accumulate, meaning overall WER can hide failures in short commands or longer spoken responses.

\textbf{Training configuration.}
All adapted runs use AdamW with batch size~1 and gradient clipping (maximum norm~1.0). Validation is performed every 10 training steps. LoRA adapters have rank $r=8$, scaling factor $\alpha=16$, and dropout 0.05. They are trained with a learning rate \texttt{1e-4} for up to 300 training steps per fold.
The base Whisper model weights are frozen; only the LoRA parameters introduced by the selected placement are updated.

\subsection{Evaluation Protocol}
\label{sec:protocol}

All experiments use leave-one-out cross-validation. For each of the 215 folds, one recording is held out for testing. From the remaining 214 recordings, one is reserved inside the fold for validation and early stopping, and the others are used for weight updates. The same fold construction is used for all model and adapter configurations. Training stops when validation loss fails to improve by $10^{-4}$ for five consecutive evaluations. Because each fold includes only one validation utterance, early stopping can be noisy; therefore, all reported deltas are computed fold-by-fold and then averaged.

\subsection{Evaluation Metrics}
\label{sec:metrics}

All metrics are computed after text normalisation: references and hypotheses are lowercased and punctuation is removed. Let $S$, $D$, and $I$ denote substitutions, deletions, and insertions, and let $N$ denote the length of the reference sequence. For word-level scoring, $N=N_w$ is the number of reference words; for character-level scoring, $N=N_c$ is the number of reference characters.
We report Word Error Rate (WER) and Character Error Rate (CER) as percentages:
\[
\mathrm{WER} = \frac{S_w + D_w + I_w}{N_w}\times 100,
\qquad
\mathrm{CER} = \frac{S_c + D_c + I_c}{N_c}\times 100.
\]
WER is the primary metric because word-level errors directly affect
command recognition and transcription usability.
CER shows whether WER changes reflect complete word failures or smaller
character-level near misses.

We also report two auxiliary similarity metrics, where higher values are better, to check whether WER/CER trends reflect broader transcript quality rather than a metric-specific effect.
The chrF score is computed from character $n$-gram precision $P$ and recall $R$ using the standard F-score form:
\[
F_{\beta} = (1 + \beta^2)\frac{P R}{\beta^2 P + R}.
\]
Cosine similarity is computed between reference ($\mathbf{r}$) and
hypothesis ($\mathbf{h}$) embeddings from a multilingual MiniLM L12 v2 \cite{wang2021minilmv2}:
\[
\cos(\mathbf{r},\mathbf{h}) =
\frac{\mathbf{r}\cdot\mathbf{h}}{\|\mathbf{r}\|_2\,\|\mathbf{h}\|_2}.
\]
chrF rewards partial character $n$-gram overlap in inflected words, while cosine similarity gives a softer sentence-level indication of semantic closeness.
They are not substitutes for WER/CER, but corroborate whether adaptation improves the transcription as a whole.

For adaptation results, deltas are measured relative to the corresponding vanilla baseline:
\[
\Delta\mathrm{WER} = \mathrm{WER}_{\mathrm{after}} -
\mathrm{WER}_{\mathrm{vanilla}},
\qquad
\Delta\mathrm{CER} = \mathrm{CER}_{\mathrm{after}} -
\mathrm{CER}_{\mathrm{vanilla}}.
\]
Negative deltas therefore indicate improvement. For the efficiency plot, WER reduction is reported with the opposite sign,
\[
\mathrm{WER\ reduction} = \mathrm{WER}_{\mathrm{vanilla}} -
\mathrm{WER}_{\mathrm{after}},
\]
so positive values indicate improvement.
All values are computed per held-out fold and then averaged across the 215 folds.

\section{Results}
\label{sec:results}

\subsection{Zero-shot Baselines}
\label{sec:results_baseline}

Table~\ref{tab:baseline_all} reports the vanilla Whisper baselines. WER and CER decrease monotonically with model size: tiny starts at 68.8/24.9, while medium reaches 12.2/5.0.

\begin{table}[h]
    \centering
    \caption{Vanilla Whisper baselines (215-fold LOO average). WER and CER are percentages.}
    \label{tab:baseline_all}
    \begin{tabular*}{0.78\textwidth}{@{\extracolsep{\fill}}lrrrr@{}}
        \toprule
        \textbf{Model} & \textbf{WER} & \textbf{CER} & \textbf{chrF} & \textbf{Cosine} \\
        \midrule
        tiny   & 68.8 & 24.9 & 61.2 & 0.600 \\
        base   & 46.2 & 15.7 & 73.5 & 0.738 \\
        small  & 23.2 &  8.6 & 87.0 & 0.885 \\
        medium & \textbf{12.2} &  \textbf{5.0} & \textbf{92.0} & \textbf{0.941} \\
        \bottomrule
    \end{tabular*}
\end{table}

The baseline results suggest that capacity alone reduces a large part of the mismatch, but not all of it.
WER drops by 22.6 percentage points from tiny to base and by 23.0 points from base to small, then by a smaller 11.0 points from small to medium.
CER follows the same direction, with the largest drop between tiny and base.
This suggests diminishing returns from scaling alone: medium is already much stronger than the smaller models, but its 12.2\% WER still leaves a clear opportunity for speaker-specific adaptation.

\subsection{Fixed-rank LoRA Comparison}
\label{sec:results_lora}

Table~\ref{tab:lora_overall} gives the main fixed-rank comparison. The table is arranged by model and strategy so that WER and CER can be read directly instead of being compressed into the same cell. Negative deltas indicate improvement relative to the corresponding vanilla baseline.
LoRA full gives the lowest observed WER and CER for every model. Decoder-only LoRA performs weakly in this protocol: it worsens base, small, and medium WER and has little CER benefit. Encoder-only LoRA helps tiny and base, but gives only marginal gains for small and worsens medium.

\begin{table}[h]
    \centering
    \caption{Overall fixed-rank LoRA comparison. WER and CER are percentages;
    $\Delta$ (pp) values are absolute changes from the vanilla baseline;
    $\Delta$ (\%) values are the corresponding relative changes.}
    \label{tab:lora_overall}
    \small
    \setlength{\tabcolsep}{3pt}
    \begin{tabular*}{0.98\textwidth}{@{\extracolsep{\fill}}llrrrrrr@{}}
        \toprule
        \textbf{Model} & \textbf{Strategy} & \multicolumn{2}{c}{\textbf{Error rate}} & \multicolumn{2}{c}{\textbf{Change (pp)}} & \multicolumn{2}{c}{\textbf{Change (\%)}} \\
        \cmidrule(lr){3-4}\cmidrule(lr){5-6}\cmidrule(lr){7-8}
        & & \textbf{WER} & \textbf{CER} & \textbf{$\Delta$ WER} & \textbf{$\Delta$ CER} & \textbf{$\Delta$ WER} & \textbf{$\Delta$ CER} \\
        \midrule
        tiny   & Vanilla      & 68.8 & 24.9 & --      & --      & --         & --         \\
               & LoRA decoder & 68.1 & 25.0 & $-0.7$  & $+0.1$  & $-1.0\%$   & $+0.4\%$   \\
               & LoRA encoder & 60.3 & 20.1 & $-8.5$  & $-4.8$  & $-12.4\%$  & $-19.3\%$  \\
               & LoRA full    & \textbf{53.8} & \textbf{17.2} & \textbf{$-15.0$} & \textbf{$-7.7$} & \textbf{$-21.8\%$} & \textbf{$-30.9\%$} \\
        \midrule
        base   & Vanilla      & 46.2 & 15.7 & --      & --      & --         & --         \\
               & LoRA decoder & 47.5 & 16.0 & $+1.3$  & $+0.3$  & $+2.8\%$   & $+1.9\%$   \\
               & LoRA encoder & 42.4 & 14.0 & $-3.8$  & $-1.7$  & $-8.2\%$   & $-10.8\%$  \\
               & LoRA full    & \textbf{35.0} & \textbf{10.8} & \textbf{$-11.2$} & \textbf{$-4.9$} & \textbf{$-24.2\%$} & \textbf{$-31.2\%$} \\
        \midrule
        small  & Vanilla      & 23.2 &  8.6 & --      & --      & --         & --         \\
               & LoRA decoder & 24.1 &  8.6 & $+0.9$  & $+0.0$  & $+3.9\%$   & $0.0\%$    \\
               & LoRA encoder & 22.9 &  8.4 & $-0.3$  & $-0.2$  & $-1.3\%$   & $-2.3\%$   \\
               & LoRA full    & \textbf{19.6} & \textbf{7.5} & \textbf{$-3.6$} & \textbf{$-1.1$} & \textbf{$-15.5\%$} & \textbf{$-12.8\%$} \\
        \midrule
        medium & Vanilla      & 12.2 &  5.0 & --      & --      & --         & --         \\
               & LoRA decoder & 12.8 &  5.2 & $+0.6$  & $+0.2$  & $+4.9\%$   & $+4.0\%$   \\
               & LoRA encoder & 12.7 &  5.1 & $+0.5$  & $+0.1$  & $+4.1\%$   & $+2.0\%$   \\
               & LoRA full    & \textbf{10.4} & \textbf{4.5} & \textbf{$-1.8$} & \textbf{$-0.5$} & \textbf{$-14.8\%$} & \textbf{$-10.0\%$} \\
        \bottomrule
    \end{tabular*}
\end{table}

Two patterns are visible in the LoRA comparison.
First, adaptation has the largest absolute payoff when the zero-shot baseline is weak: LoRA full reduces WER by 15.0 percentage points for tiny and 11.2 points for base, compared with 3.6 and 1.8 points for small and medium. In relative terms, however, the gap across model sizes is considerably narrower: LoRA full achieves $-21.8\%$, $-24.2\%$, $-15.5\%$, and $-14.8\%$ relative WER reduction for tiny, base, small, and medium respectively, indicating that the apparent diminishing returns in the absolute view are partly an artefact of stronger baselines having less absolute WER to reduce.
Second, the useful single-side effect is mostly on the encoder side. Decoder-only LoRA is not consistently reliable in this protocol, which suggests that the main pathological-speaker mismatch is unlikely to be fully addressed by adjusting the output-side language-generation path alone. Full-attention LoRA still improves over encoder-only LoRA, indicating that adapting decoder attention becomes useful when it is paired with an adapted acoustic representation.

Table~\ref{tab:lora_after} reports the post-adaptation chrF and cosine scores, avoiding repetition of the WER/CER values in Table~\ref{tab:lora_overall}.

\begin{table}[h]
    \centering
    \caption{Post-adaptation chrF and cosine for the three LoRA placements.}
    \label{tab:lora_after}
    \begin{tabular*}{0.82\textwidth}{@{\extracolsep{\fill}}llrr@{}}
        \toprule
        \textbf{Model} & \textbf{Strategy} & \textbf{chrF} & \textbf{Cosine} \\
        \midrule
        tiny   & LoRA decoder & 61.1 & 0.611 \\
               & LoRA encoder & 66.8 & 0.654 \\
               & LoRA full    & \textbf{71.2} & \textbf{0.701} \\
        \midrule
        base   & LoRA decoder & 73.8 & 0.747 \\
               & LoRA encoder & 77.4 & 0.773 \\
               & LoRA full    & \textbf{81.3} & \textbf{0.814} \\
        \midrule
        small  & LoRA decoder & 85.9 & 0.870 \\
               & LoRA encoder & 86.7 & 0.886 \\
               & LoRA full    & \textbf{88.7} & \textbf{0.894} \\
        \midrule
        medium & LoRA decoder & 91.6 & 0.935 \\
               & LoRA encoder & 92.0 & 0.936 \\
               & LoRA full    & \textbf{92.8} & \textbf{0.946} \\
        \bottomrule
    \end{tabular*}
\end{table}

The chrF and cosine scores support the WER/CER pattern rather than showing a metric-specific trade-off.
For every model size, LoRA full gives the best similarity scores as well as the best error rates.
The margin is largest for tiny and base, where the acoustic mismatch is most severe, and becomes smaller for small and medium. This consistency across word errors, character errors, character $n$-gram overlap, and sentence-level cosine similarity strengthens the interpretation that full-attention LoRA improves the transcription as a whole, not only a single metric.

\FloatBarrier
\subsection{Training-efficiency Trade-off}
\label{sec:results_efficiency}

\begin{figure}[t]
    \centering
    \includegraphics[width=\textwidth]{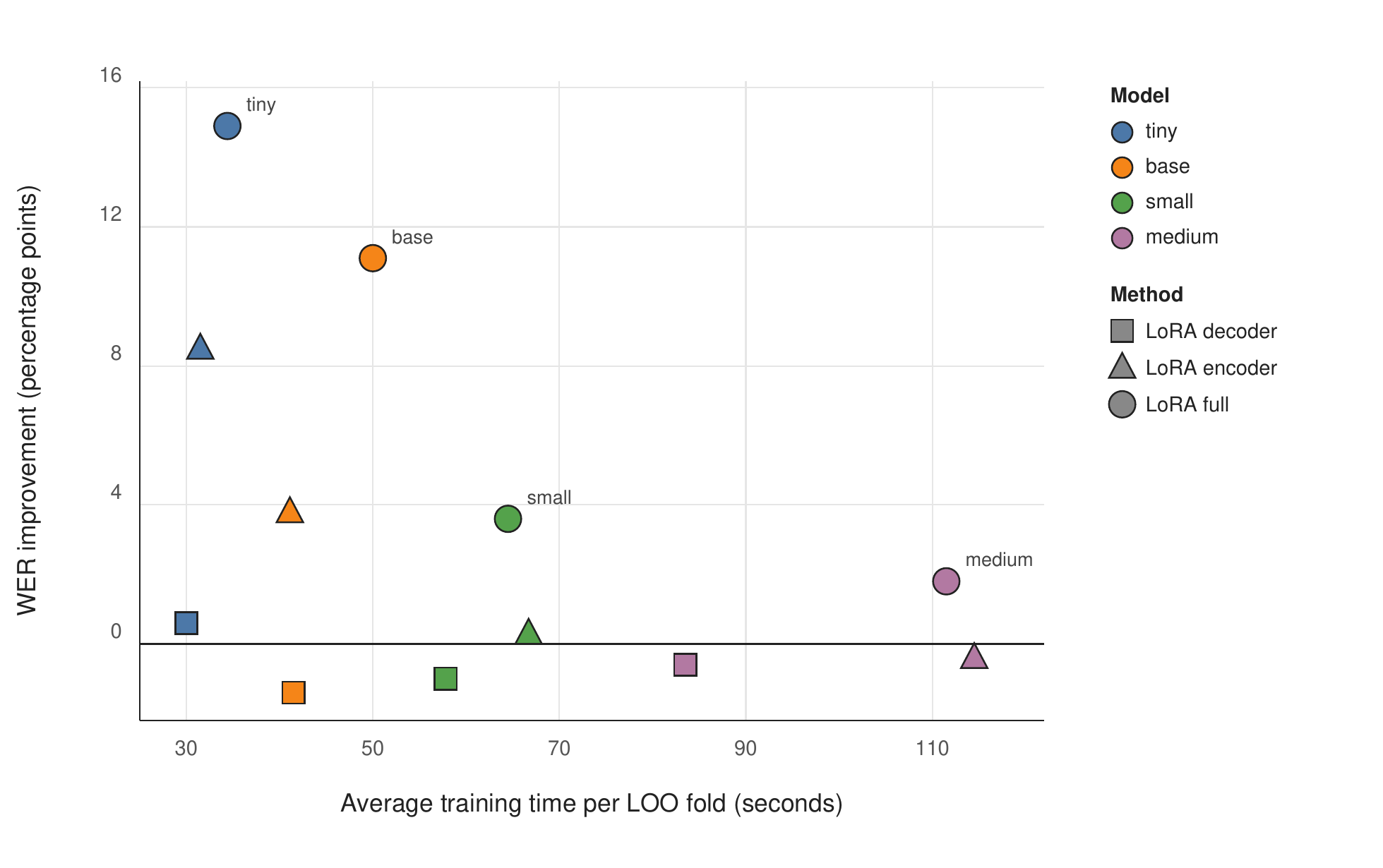}
    \caption{Training time versus WER improvement for fixed-rank LoRA configurations. Each point corresponds to one model size and one LoRA
    placement.}
    \label{fig:time_wer}
\end{figure}

Figure~\ref{fig:time_wer} compares the average wall-clock training time per leave-one-out fold with the corresponding WER reduction.
Timing is measured consistently for the LoRA update stage and averaged across folds, so it should be read as relative within-run efficiency. Each point is one fixed-rank LoRA configuration. The x-axis reports average training time per fold, and the y-axis reports the absolute WER reduction in percentage points; negative values indicate that adaptation worsened WER.

In these runs, LoRA full provides the most favourable observed time--accuracy trade-off across the considered model sizes.
For tiny and base, it gives large WER reductions of 15.0 and 11.2 percentage points with average fold times of 34.4~s and 50.0~s, respectively. For small and medium, the gains are smaller, 3.6 and 1.8 percentage points, but LoRA full remains the most effective configuration. Decoder-only adaptation is not efficient in this protocol: it often gives little improvement, or even degrades WER, despite similar training time.
Thus the extra attention-side coverage of LoRA full is justified for the smaller models by a much larger gain, while for medium it is mainly a choice for maximising accuracy when even a small WER reduction matters.

\FloatBarrier
\subsection{Duration-tier Analysis}
\label{sec:results_tier}

Table~\ref{tab:tier_full} reports the duration-tier behaviour for the lowest-WER LoRA placement, LoRA full, including the corresponding after-minus-before deltas.
It uses one row per model--tier pair to avoid compressing WER and CER into long multi-metric cells.
The main weakness is small-long recordings, where LoRA full worsens both WER and CER.

\begin{table}[!h]
    \centering
    \caption{LoRA full performance and deltas by duration tier. WER and CER are percentages; $\Delta$ pp values are after minus before in percentage points; $\Delta$ \% values are the corresponding relative changes. Negative deltas indicate error reduction; bold deltas mark the largest reduction within each tier metric.}
    \label{tab:tier_full}
    \scriptsize
    \setlength{\tabcolsep}{3pt}
    \renewcommand{\arraystretch}{0.88}
    \begin{tabular*}{0.98\textwidth}{@{\extracolsep{\fill}}llrrrrrrrr@{}}
        \toprule
        \textbf{Model} & \textbf{Tier} & \multicolumn{4}{c}{\textbf{WER}} & \multicolumn{4}{c}{\textbf{CER}} \\
        \cmidrule(lr){3-6}\cmidrule(lr){7-10}
        & & \textbf{Before} & \textbf{After} & \textbf{$\Delta$ pp} & \textbf{$\Delta$ \%} & \textbf{Before} & \textbf{After} & \textbf{$\Delta$ pp} & \textbf{$\Delta$ \%} \\
        \midrule
        tiny   & Short  & 80.1 & 71.8 & $-8.3$  & $-10.4\%$  & 46.0 & 39.4 & $-6.6$  & $-14.3\%$  \\
               & Medium & 71.6 & 51.0 & \textbf{$-20.6$} & \textbf{$-28.8\%$} & 24.4 & 14.5 & \textbf{$-9.9$} & \textbf{$-40.6\%$} \\
               & Long   & 57.2 & 54.1 & $-3.1$  & $-5.4\%$   & 18.2 & 15.4 & $-2.8$  & $-15.4\%$  \\
        \midrule
        base   & Short  & 74.8 & 52.2 & \textbf{$-22.6$} & \textbf{$-30.2\%$} & 41.1 & 33.1 & \textbf{$-8.0$} & \textbf{$-19.5\%$} \\
               & Medium & 43.8 & 34.8 & $-9.0$  & $-20.5\%$  & 13.6 & 8.6  & $-5.0$  & $-36.8\%$  \\
               & Long   & 41.1 & 29.1 & \textbf{$-12.0$} & \textbf{$-29.2\%$} & 11.4 & 7.9  & \textbf{$-3.5$} & \textbf{$-30.7\%$} \\
        \midrule
        small  & Short  & 45.4 & 35.8 & $-9.6$  & $-21.1\%$  & 30.3 & 27.0 & $-3.3$  & $-10.9\%$  \\
               & Medium & 21.4 & 16.7 & $-4.7$  & $-22.0\%$  & 6.6  & 4.7  & $-1.9$  & $-28.8\%$  \\
               & Long   & 19.0 & 20.8 & $+1.8$  & $+9.5\%$   & 5.3  & 7.4  & $+2.1$  & $+39.6\%$  \\
        \midrule
        medium & Short  & 31.4 & 31.4 & $0.0$   & $0.0\%$    & 28.4 & 27.4 & $-1.0$  & $-3.5\%$   \\
               & Medium & 9.2  & 7.3  & $-1.9$  & $-20.7\%$  & 2.2  & 1.7  & $-0.5$  & $-22.7\%$  \\
               & Long   & 12.5 & 10.3 & $-2.2$  & $-17.6\%$  & 3.3  & 2.9  & $-0.4$  & $-12.1\%$  \\
        \bottomrule
    \end{tabular*}
\end{table}

The duration-tier results indicate that personalisation is not uniform
across utterance length.
Medium utterances improve for all models ($\Delta$WER from $-1.9$~pp to $-20.6$~pp
), while short utterances remain difficult, with medium unchanged
at 31.4\% WER.
Long utterances improve except for small, where WER/CER degrade by
+1.8/+2.1 percentage points; the 21-recording short tier may also be
fold-sensitive.

\FloatBarrier

\section{Conclusion}
\label{sec:conclusion}

In this single-speaker case study, we evaluate fixed-rank LoRA personalisation of Whisper for an Italian speaker with pathological speech under data-scarce conditions. Larger models achieve lower error rates, but adaptation remains highly sensitive to placement. In particular, decoder-only LoRA is often ineffective or detrimental, whereas full-attention LoRA provides the most consistent placement. These results suggest that speaker-specific mismatch in pathological speech is better addressed by adapting both acoustic and decoder attention, rather than modifying the output pathway alone. Scaling reduces the absolute WER payoff of LoRA full---from 15.0~pp for tiny to 1.8~pp for medium---but in relative terms the reduction remains within a narrower 15--24\% band across all four model sizes, indicating that the apparent diminishing returns are partly an artefact of larger models starting from a stronger zero-shot baseline.

Across the analyses, the same pattern is observed. Full-attention LoRA consistently achieves the most favourable time--accuracy trade-off, and auxiliary metrics (CER, chrF, cosine similarity) follow the same trends as WER. Duration-tier analysis further shows that performance gains are not uniform across utterance lengths: medium handles long utterances well but struggles on short ones, while the small model degrades on long utterances under full-attention LoRA. 

\paragraph{Limitations.}
As this is a single-speaker case study, the findings should not be interpreted as population-level results; future work should evaluate broader and more diverse cohorts across speakers, pathology types, severity levels, languages, and recording conditions, and investigate whether adapting or controlled fine-tuning at the currently frozen pre-attention encoder front-end can further reduce the remaining performance gap. Several further limitations deserve acknowledgement. First, because recording conditions and detailed pathology metadata were unavailable, the acoustic deviations reported in Section~\ref{sec:acoustic_characterisation} cannot be fully disentangled from possible recording-condition effects; controlled recordings with documented conditions and a matched healthy-speech baseline would be needed to isolate the pathology-specific contribution. Second, the speaker's specific pathology and severity could not be disclosed due to ethical and privacy constraints on the dataset, which limits the interpretability of the acoustic characterisation. Third, both absolute (pp) and relative (\%) WER and CER reductions are now reported in Tables~\ref{tab:lora_overall} and~\ref{tab:tier_full} for transparency; however, the analysis does not include formal statistical significance testing, and---given the variability of single-utterance folds under leave-one-out cross-validation---paired tests across folds would further strengthen confidence in the reported differences. Finally, the study is restricted to LoRA, and comparing against other parameter-efficient methods already surveyed in Related Work---such as BitFit~\cite{benzaken2022bitfit} and adapter-based approaches~\cite{tomanek2021residual}---would clarify whether the observed placement effects are specific to low-rank adaptation or generalise across PEFT strategies.

\begin{credits}

\subsubsection{\ackname}
This work was partially supported by computational resources provided by CINECA. We acknowledge the CINECA award under the ISCRA initiative for the availability of high-performance computing resources and support.

\end{credits}

\bibliographystyle{splncs04}
\bibliography{references}

\end{document}